# Coupling spin defects in hexagonal boron nitride to a microwave cavity

**Thinh N. Tran[1], Angus Gale[2], Benjamin Whitefield[1,2], Milos Toth[1,2], Igor Aharonovich[1,2] and Mehran Kianinia[1,*]**

[1] School of Mathematical and Physical Sciences, Faculty of Science, University of Technology Sydney, Ultimo, New South Wales 2007, Australia
[2] ARC Centre of Excellence for Transformative Meta-Optical Systems (TMOS), University of Technology Sydney, Ultimo, New South Wales 2007, Australia

E-mail: mehran.kianinia@uts.edu.au

**Abstract**
Optically addressable spin defects in hexagonal boron nitride (hBN) have become a promising platform for quantum sensing. While sensitivity of these defects are limited by their interactions with the spin environment in hBN, inefficient microwave delivery can further reduce their sensitivity. Hare, we design and fabricate a microwave double arc resonator for efficient transferring of the microwave field at 3.8 GHz. The spin transitions in the ground state of $V_B^-$ are coupled to the frequency of the microwave cavity which results in enhanced optically detected magnetic resonance (ODMR) contrast. In addition, the linewidth of the ODMR signal further reduces, achieving a magnetic field sensitivity as low as 42.4 µT/√Hz. Our robust and scalable device engineering is promising for future employment of spin defects in hBN for quantum sensing.

## 1. Introduction

Optically active spin defects constitute the main quantum hardware for applications in sensing and communication technologies[1-5]. Among existing solid state materials hexagonal boron nitride, a wide bang gap two dimensional material, has been shown to host variety of spin defects at room temperature[6-8]. Recently, a new class of spin defects - namely the negatively charged boron vacancy $V_B^-$ defects in hexagonal boron nitride (hBN) has emerged as a promising candidate for quantum sensing[6, 9]. The $V_B^-$ emits at ~ 810 nm, and has a ground state spin transition at ~ 3.5 GHz. Coherent control of the spin state as well as preliminary imaging experiments were demonstrated, as a proof of principle to utilize this defect as a quantum sensor[9-15]. Furthermore, the superior properties of layered materials in achieving the precise thickness of host hBN and hence controlling the distance between the quantum sensor and the sample has brought much attention and excitement to the community with the possibility of performing quantum sensing at these unexplored regimes. With these fundamental attributes, the $V_B^-$ defects have the potential to become an important tool to study physical properties of emerging 2D materials, devices and heterostructures[9, 10, 16].

To deliver the microwave fields, necessary to control and manipulate the spin state, metal waveguides such as gold stripes are commonly used[12, 13, 17]. While microwave delivery can be easily achieved by transferring a hBN flake on top of a metal stripe, it is an inefficient way to deliver the microwaves with losses. In an alternative approach, one can design and engineer a microwave cavity that resonates with the transitions of the defect spin states[18-23]. Such a cavity is posed to enhance ODMR contrast and prevent microwave power broadening, thus enhancing the spin sensitivity. In turn, the improvement of ODMR contrast and sensitivity of $V_B^-$ defects in hBN is highly sought after for magnetic, thermal and pressure sensing applications.

In this work, we effectively facilitate the process of designing and fabricating a microwave resonators that matches to the upper transition of $V_B^-$, from $|+1\rangle$ to $|0\rangle$ in the ground level. The microwave resonators were engineered by low cost printed circuit boards (PCBs) which can easily be tuned by changing the inner arc radius. Our results show the improvement in ODMR contrast and the magnetic field detection sensitivity of the hybrid structure, paving the way to integrating a microwave cavity to spin defects in hBN for ultra-high sensitive quantum sensing.

To create the $V_B^-$ defects, we have used a low energy nitrogen ion beam at 30 KeV to generate as shown in figure 1a. First, hBN flakes were exfoliated on a clean silicon substrate with a thin layer of thermal oxide and further cleaned in a UV ozone chamber for 15 minutes to remove any organic residuals from the surface. During the implantation, the nitrogen ion beam was maintained at $2 \times 10^{16}$ ion·cm$^{-2}$ with ion current at 21.9 pA[24, 25].

The $V_B^-$ defects are a spin 1 system with a triplet at ground state separated by ~3.47 GHz between m$_s$ = 0 and m$_s$ = ±1 states, as shown schematically in figure 1b. The degeneracy of the latter states are lifted even at zero external magnetic field. This is evident with the two distinct resonances, $v_{1,2}$ in the ODMR spectra of the $V_B^-$ defect. The resonant frequencies, $v_{1,2}$ are generalized under external magnetic field (B) as $v_{1,2} = D_{gs}/h + (1/h)\sqrt{E_{gs}^2 + (g\mu_B B)^2}$, where $D_{gs}$ and $E_{gs}$ are zero-field splitting parameters, $g$ is the Landé factor, $\mu_B$ is the Bohr magneton and $h$ is Planck's constant. Without an external magnetic field, the two resonant frequencies only split about $E_{gs}/h \approx 50$ MHz which could adversely affect the characterization of the microwave resonator and the $V_B^-$ defects. Therefore, we intentionally designed the resonant frequency of the resonator, $\omega_c$ at ≈ 3.8 GHz and tune the upper resonant frequency, $v_2$ to match $\omega_c$ with an external magnetic field (Figure 1b).

To confirm the successful generation of the $V_B^-$ defects a confocal microscopy characterization was carried out to detect the photoluminescence (PL) emission of the defects centered at ~800 nm (figure 1c). Next, the hBN flake was transferred on to the fabricated microwave cavity which resonates with the transition between m$_s$ = +1 and m$_s$ = 0 in the ground state of $V_B^-$.

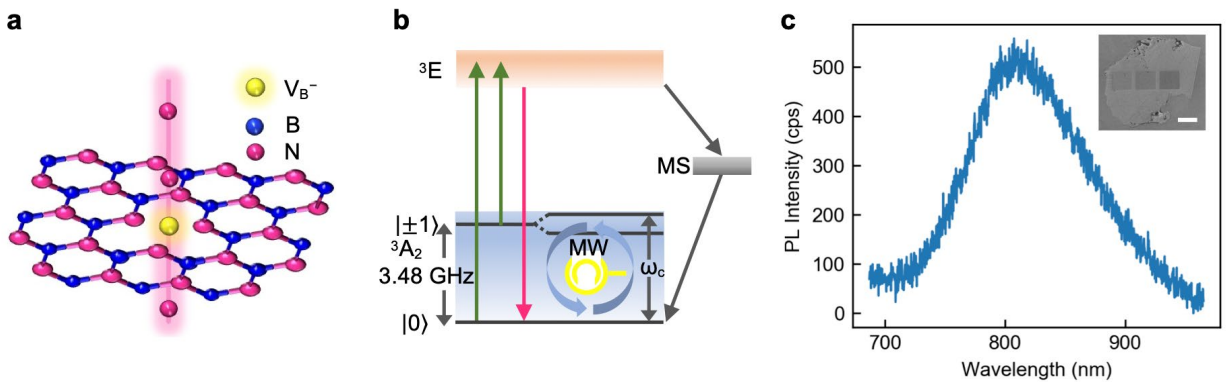

**Figure 1.** Spin defects in hBN. a) Creation of $V_B^-$ by nitrogen (N) ion beam implantation into the hBN lattice. b) Electronic level structure of $V_B^-$ in hBN with zero-field splitting at ground state E$_{gs}$ = 3.48 GHz. By using an external magnetic field, the ground state spin triplet is split into transitions between m$_s$ = +1 and m$_s$ = 0 to match the resonance frequency of the microwave cavity ($\omega_c$). c) Photoluminesce spectrum from representative $V_B^-$ defects in hBN. Inset, the scanning electron microscope (SEM) of a hBN flake after N ion beam radiation on areas of 50 × 50 μm$^2$ (scale bar: 50 μm).

Figure 2a shows the double arc resonator on a PCB (Rogers 4350B substrate) with a compact size (~ 28 × 15 mm$^2$). The resonator consists of two closely separated arcs by a small gap. The microwave signal is delivered through a standard 50 Ω microstrip line which forms a capacitor with the double gap-arc by a small distance. The design was inspired by the proposal from Shamonin et al[18]. To further characterize the design and fine tune resonant frequency, the electromagnetic numerical simulations (CST Studio Suite) were used. Figure 2b shows the simulated magnetic field strength distribution at the frequency of 3.78 GHz without any loads. The magnetic strength is uniform at the center of the double gap-arc, however it concentrates along the rims of the outer arc.

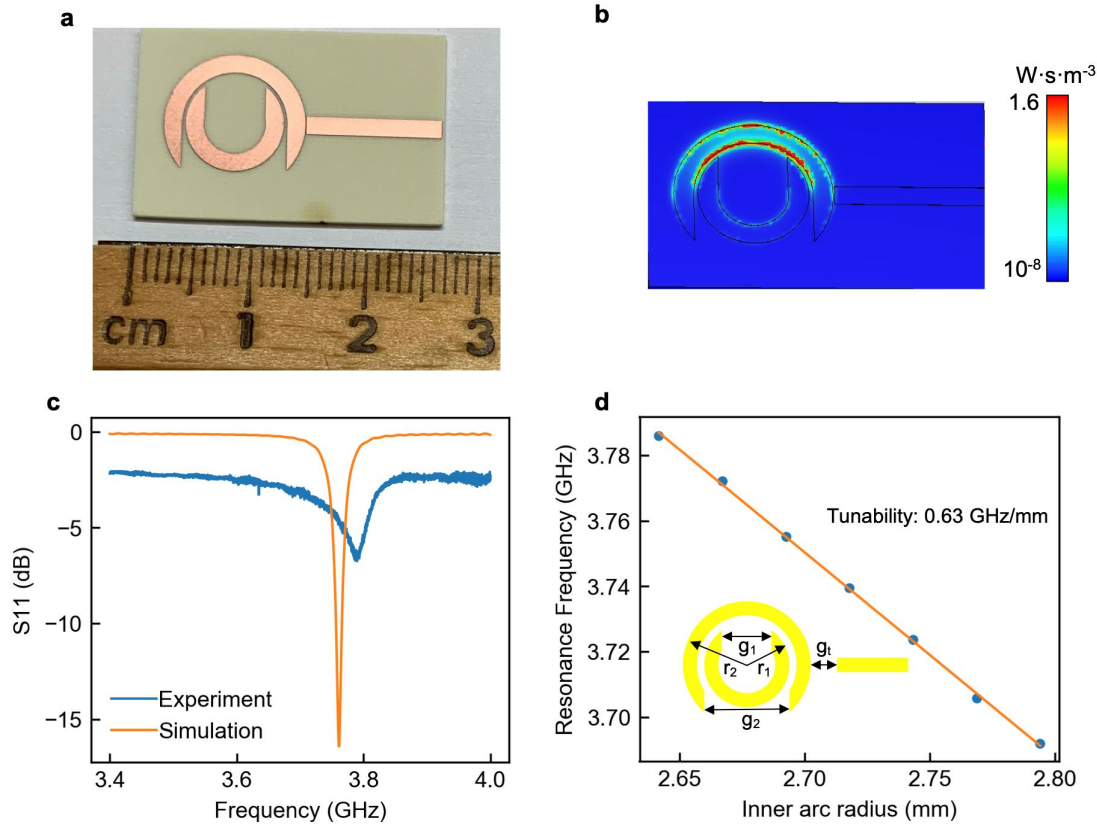

**Figure 2.** Microwave resonator characterizations. a) A double arc resonator on a PCB. b) Simulated magnetic field strength distribution of the resonator with the intensity color bar on the right. c) Simulated (orange) and measured (blue) return loss (S11) of a resonator with geometrical parameters listed in the text. d) Resonance frequency of the double arc resonator as a function of the radius of the inner arc together with the 2D design of the resonator. $r_1$ and $r_2$ are the radii of the inner and outer arcs, respectively; $g_1$ and $g_2$ are the cut width of the inner and outer arcs, respectively; $g_t$ is the distance between the transmission line to the outer arc.

The resonant frequency of the resonator was calculated from the simulated results of return loss (S$_{11}$) of the resonator as shown in Figure 2c, and confirmed experimentally by measuring the transmission of the resonator. The resonance has a marginally lower Q factor of ~ 65 and resonant frequency ~ 3.8 GHz. These differences are due to the capacitive coupling modified by a metal holder used to mount the resonator to a piezo scanning stage. The design parameters of the resonator given in Table 1.

The resonance frequency of the resonators can be tuned by modifying the radius of the inner arc $r_1$. Seven resonators with different inner radius were fabricated and measured to identify the resonance frequency. The correlation between the resonance frequencies and the inner radii is shown in Figure 2d. This result shows a linear dependence between the resonance frequency and the inner radius with a slope of −0.63 GHz/mm.

**Table 1.** Design parameters for the resonator with Q factor of ~ 90 and resonant frequency of 3.78 GHz

| Parameters | | | Comments |
|---|---|---|---|
| $\varepsilon_r$ ~ 3.48 | $r_1$ ~ 2.79 mm | $g_t$ ~ 101 μm | $\varepsilon_r$, $t_{Cu}$, tan($\sigma$), $w$, $g_t$, $r_1$, $r_2$, $g_1$, and $g_2$ are dielectric constant of the PCB, thickness of the copper layer on the PCB, loss tangent of the PCB material, width of copper traces, inner radius, outer radius, coupling gap, gap between the rings, and cut width of the inner and outer arcs, respectively. |
| $t_{Cu}$ ~ 89 μm | $r_2$ ~ 4.55 mm | $g_2$ ~ 9.1 mm | |
| tan($\sigma$) ~ 0.0037 | $g_1$ ~ 5.13 mm | $w$ ~ 1.6 mm | |

To couple the $V_B^-$ defects to the resonators, the hBN flake containing the defects was transferred directly onto the PCB, in the region where the microwave field is homogeneous. To shift the upper resonant frequency, $v_2$ of $V_B^-$ defects to the cavity mode, a small magnetic field applied perpendicularly to the flake. Figure 3a shows ODMR spectra of $V_B^-$ in the absence of an external magnetic field (blue) and when the signal is brought to the resonance of the microwave cavity under ~ 10mT of external magnetic field. The enhancement of ODMR contrast is increasing to ~ 7% employing the cavity resonance, compared to a pristine contrast of ~ 1.7 %, under the same microwave power (15 dBm).

To further corroborate the coupling strength of $V_B^-$ defects into the microwave resonator, we perform ODMR measurements under various microwave powers. Figure 3(b, c) show the ODMR contrast and linewidth at different microwave powers. All experiments were carried out under laser excitation with power of 2 mW. The ODMR contrast increases significantly when the B field is applied to tune the ODMR resonance to the cavity mode. Notably, even under very low microwave powers (~ -20 dBM), where the signal is non detectable under zero magnetic field, the detection becomes feasible if the ODMR signal is enhanced by the microwave cavity. At maximum, 15 dBm microwave power, the ODMR contrast increases about 3.5 times, accompanied by ~ 20% linewidth reduction, when the magnetic field is applied.

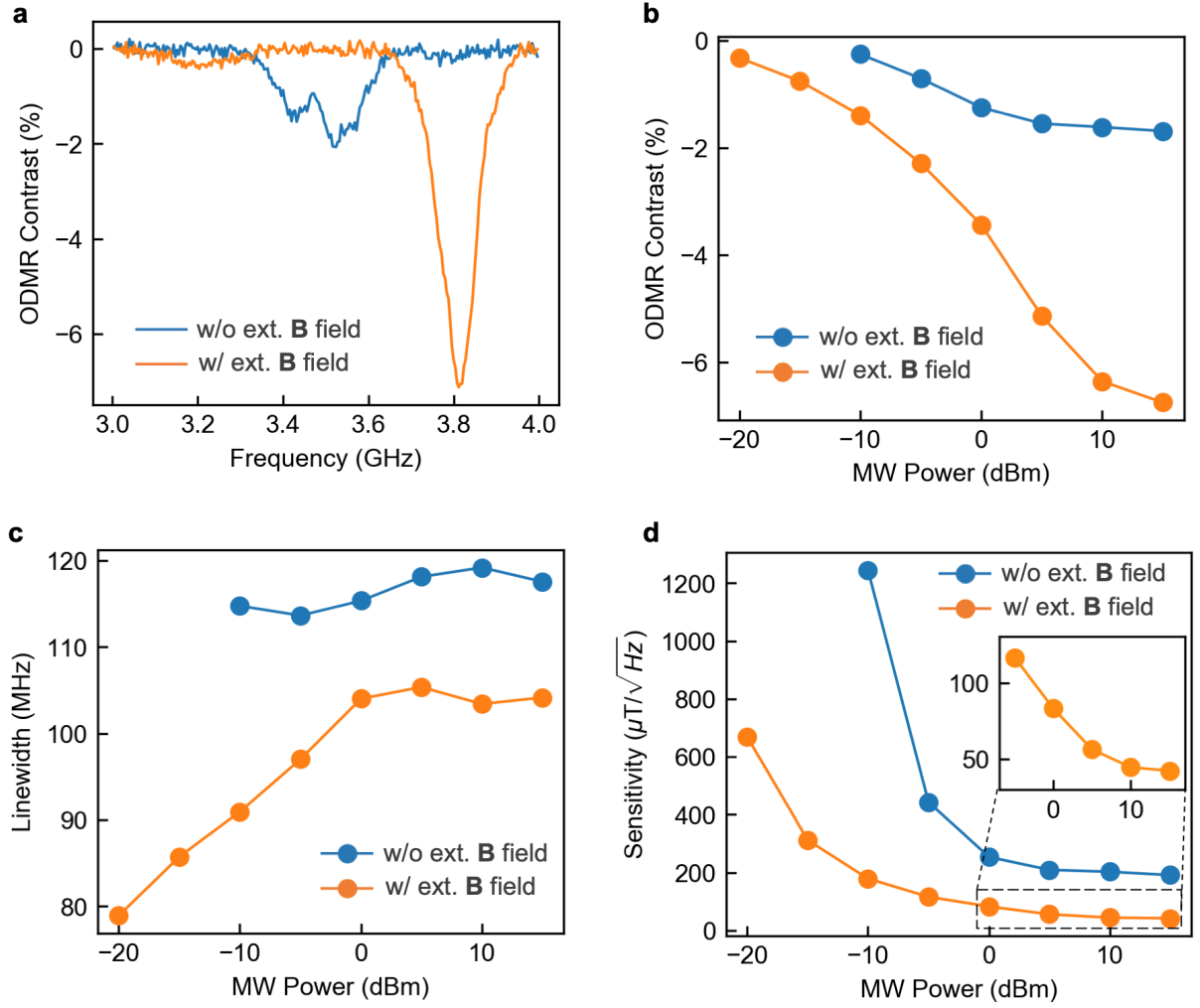

**Figure 3.** ODMR comparison between with (orange) and without (blue) an external magnetic field. a) ODMR spectrum with and without an external magnetic field at the same microwave power. b) ODMR contrasts, c) Linewidths, and d) Magnetic field sensitivity at different microwave powers with the inset showing zoom-in the sensitivity from 0 to 15 dBm when an external magnetic field is applied.

With the enhancement of ODMR contract and reduction of linewidth broadening under high microwave power, the magnetic field sensitivity is expected to be reduced. The magnetic field sensitivity is defined by the ODMR contrast $C$, average photon count rate $R$ and the linewidth $\Delta v$ as the following equation

$$\eta_B \approx P_F \frac{h}{g\mu_B} \frac{\Delta v}{C\sqrt{R}}$$

where $P_F$ is a numerical parameter related to lineshape profile. In our case, $P_F \approx 0.7$ for a Gaussian profile. Given the measured photon count rates, ODMR contrasts and linewidths, the magnetic sensitivities can be calculated as shown in Figure 3d. We observed an enhancement of about ~ 5 in magnetic sensitivity, that reaches ~42.4 µT/√Hz with the microwave resonator.

To summarize, we have demonstrated the coupling of spin defects ($V_B^-$) in hBN to a microwave resonator. The higher ODMR contrast (~ 6.8 %) and narrower linewidth (~ 104 MHz) were achieved due to the coupling between the microwave resonator and the upper resonant frequency of the $V_B^-$ under external magnetic field. Furthermore, the detectable magnetic field sensitivity can be reduced to as low as 42.2 µT/√Hz, making it appropriate for detection of small magnetic fields. This result can be further improved by using different ion irradiation schemes with the possibility of reaching the sensitivity in the range of nT/√Hz of $V_B^-$ in hBN. This improvement of sensitivity would pave the way for quantum sensing applications by using spin defects in layered hBN material.

**Acknowledgments**

This work is supported by the Australian Research Council (CE200100010, FT220100053) and the Office of Naval Research Global (N62909-22-1-2028).